\documentclass[aps, reprint, prx, superscriptaddress]{revtex4-1}
\usepackage{dcolumn}%
\usepackage{epsfig, graphics, amssymb, amsmath, color, bm}
\usepackage{epstopdf}

\usepackage{color}
\usepackage{amsmath}
\usepackage{color}
\usepackage{siunitx}
\usepackage{float}
\usepackage{soul}
\usepackage{pifont,wasysym}

\usepackage{lipsum} 

\begin{document}

\title{Self-learning how to swim at low Reynolds number}

\author{Alan Cheng Hou Tsang}
\affiliation{Department of Bioengineering, Stanford University, Stanford, CA 94305}

\author{Pun Wai Tong}
\affiliation{Clinical Genomics Program, Stanford Health Care, Stanford, CA 94305}

\author{Shreyes Nallan}
\affiliation{Department of Mechanical Engineering,
Santa Clara University, Santa Clara, CA 95053, USA}

\author{On Shun Pak}
\affiliation{Department of Mechanical Engineering,
Santa Clara University, Santa Clara, CA 95053, USA}

\date{\today}

\begin{abstract}
Synthetic microswimmers show great promise in biomedical applications such as drug delivery and microsurgery. Their locomotion, however, is subject to stringent constraints due to the dominance of viscous over inertial forces at low Reynolds number (Re) in the microscopic world. Furthermore, locomotory gaits designed for one medium may become ineffective in a different medium. Successful biomedical applications of synthetic microswimmers rely on their ability to traverse biological environments with vastly different properties. Here we leverage the prowess of machine learning to present an alternative approach to designing low Re swimmers. Instead of specifying any locomotory gaits \textit{a priori}, here a swimmer develops its own propulsion strategy based on its interactions with the surrounding medium via reinforcement learning. This self-learning capability enables the swimmer to modify its propulsion strategy in response to different environments. We illustrate this new approach using a minimal example that integrates a standard reinforcement learning algorithm ($Q$-learning) into the locomotion of a swimmer consisting of an assembly of spheres connected by extensible rods. We showcase theoretically that this first self-learning swimmer can recover a previously known propulsion strategy without prior knowledge in low Re locomotion, identify more effective locomotory gaits when the number of spheres increases, and adapt its locomotory gaits in different media. These results represent initial steps towards the design of a new class of self-learning, adaptive (or ``smart'') swimmers with robust locomotive capabilities to traverse complex biological environments.
\end{abstract}

\maketitle

\section{Introduction}

Swimming at the microscale encounters stringent constraints due to the dominance of viscous over inertial forces at low Reynolds numbers (Re) \cite{purcell1977life, lauga2009hydrodynamics}. As a result of kinematic reversibility, Purcell's scallop theorem rules out reciprocal motion (i.e., strokes with time-reversal symmetry) for effective locomotion in the absence of inertia \cite{purcell1977life}. Common macroscopic propulsion strategies thus become ineffective in the microscopic world. Microorganisms have evolved diverse locomotion strategies \cite{Fauci2006, lauga2009hydrodynamics}, for instance, by rotating helical slender appendages (termed flagella) or propagating deformation waves along flagella via actions of molecular motors, to escape the constraints of the scallop theorem. Extensive efforts in the past few decades have sought to elucidate physical principles that underlie cell motility \cite{Yeomans2014, Elgeti2015, Lauga16, Saintillan18}. This has improved our general understanding of locomotion at low Re, which in recent years has engendered a variety of synthetic microswimmers \cite{Ebbens2010, Sengupta, Hu18}.

Synthetic microswimmers capable of navigating biological environments offer exciting opportunities for biomedical applications, such as microsurgery and targeted drug delivery \cite{Nelson2010, Gao2014nano}. Purcell pioneered the design of synthetic microswimmers by inventing a sequence of movements with a three-link swimmer (known as Purcell's swimmer) in a non-reciprocal manner to generate self-propulsion \cite{purcell1977life, becker_koehler_stone_2003, Tam2007}. Subsequent interdisciplinary efforts have recently resulted in major advances in the design and fabrication of synthetic microswimmers. While some designs are bio-mimetic or bio-inspired (e.g.~swimmers with appendages that resemble flagella of microorganisms \cite{Dreyfus2005, Abbott2009, PakSoft, Zhang2009, Ghosh2009}), others ingeniously exploit physical (e.g.~Najafi-Golestanian's swimmer \cite{Najafi2004} and Purcell's `rotator' \cite{Dreyfus2005B}) and/or physico-chemical (e.g.~catalytic Janus motors \cite{Paxton2006, Moran2017}) mechanisms available in the microscopic world to self-propel in the absence of inertia.

Successful biomedical applications of synthetic microswimmers rely on their ability to traverse vastly different biological environments, including blood-brain, gastric mucosal barriers, and tumor micro-environments \cite{Nassif2002, Celli2009, Mirbagheri2016}. Despite significant progress over the past decades, existing microswimmers are typically designed to have fixed locomotory gaits for a particular type of medium or environmental condition. However, gaits that are optimal in one medium may become ineffective in a different medium; hence, locomotion performance of synthetic microswimmers with fixed locomotory gaits may not be robust to environmental changes. In contrast, natural organisms show robust locomotion performance across varying environments by adapting their locomotory gaits to the surroundings \cite{BARCLAY1946, Christopher2010, Maladen2009}. Without adaptability like their biological counterparts, it remains formidable for synthetic microswimmers to operate in complex biological media with unpredictable environmental factors. Novel approaches via modular microrobotics and the use of soft active materials have been recently proposed to tackle these challenges \cite{Cheang2016, Hu2018}.

Here we leverage the prowess of machine learning to investigate a new approach in designing low Re swimmers. Machine learning has enabled the design of artificial intelligent systems that can perform complex tasks without being explicitly programmed \cite{jordan2015}. This approach has also sparked several novel directions in fluid mechanics, including modeling of turbulence \cite{ling_kurzawski_templeton_2016, kutz_2017}, fish schooling \cite{Gazzola2014, Gazzola2016, Verma201800923}, soaring birds \cite{ReddyE4877}, wake detection \cite{Colvert2018}, and navigation problems \cite{Colabrese2017, Landin}. Here we ask the general questions: Without any prior knowledge on low Re locomotion, can a swimmer learn how to escape the constraints of the scallop theorem for self-propulsion via a simple machine learning algorithm? How well does this self-learning approach perform for a system with multiple degrees of freedom? Can such a self-learning swimmer adapt its locomotory gaits to traverse media with vastly different properties?

\begin{figure*}[t]
\centering
\includegraphics[width=1\textwidth]{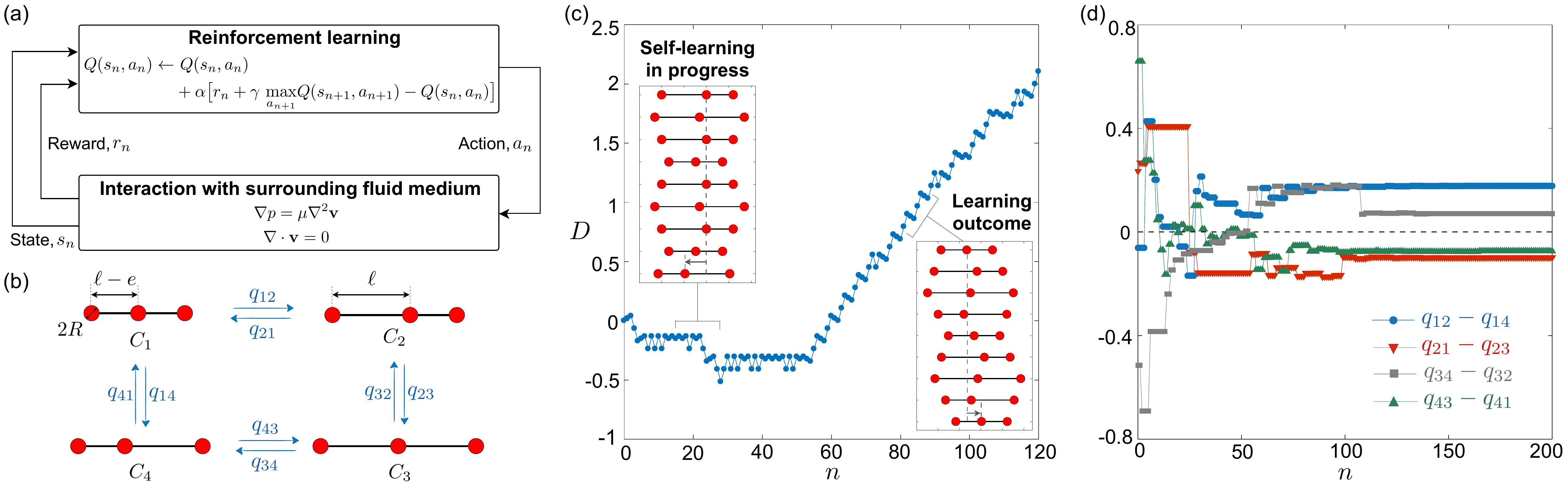}
\vspace{-0.3cm}
\caption{Reinforcement learning enables a swimmer to self-learn at low Re. (a) Schematic of reinforcement learning of a swimmer that progressively learns how to swim by interacting with the surrounding fluids. (b) A state diagram for a three-sphere swimmer. $q_{ij}$ are entries in the $Q$-matrix that evolve based on reinforcement learning. (c) A typical learning process of a self-learning three-sphere swimmer. The dimensionless cumulative displacement $D$ of the swimmer evolves over learning steps $n$. The swimmer initially struggles to propel (left inset). However, after accumulating sufficient knowledge (e.g., $n>60$), the swimmer develops an effective propulsion policy (right inset) that repeats the same sequence of actions over time (except for small fluctuations in $D$ due to the $\epsilon$ probability of random selection of actions). The learning outcome is consistent with Najafi-Golestanian's swimmer \cite{Najafi2004}. (d) The evolution of the differences of entries in the $Q$-matrix for the case shown in (c). In (c) \& (d), $\gamma=0.8$, $\epsilon=0.05$.} 
\label{fig:learning}
\end{figure*}

In this work we present the first example of integrating machine learning into the design of locomtory gaits at low Re: instead of specifying a locomotory gait \textit{a priori}, the swimmer develops its own propulsion policy based on its interactions with the surrounding medium. As a first step, we demonstrate the potential power of this machine learning approach via a simple reconfigurable system (an assembly of spherical particles) with a standard reinforcement learning framework ($Q$-learning)  [Fig.~\ref{fig:learning}(a)]. Specifically, we show that, without requiring any prior knowledge on low Re locomotion, a self-learning swimmer can recover the swimming strategy by Najafi and Golestanian \cite{Najafi2004}, identify more effective locomotory gaits with increased degrees of freedom, and adapt locomotory gaits in different media. These results represent initial steps towards the design of a new class of self-learning, adaptive (or ``smart'') swimmers with robust locomotive capabilities to traverse complex biological environments.

\section{Swimming at low Re via reinforcement learning} 

We considered a simple reconfigurable system for locomotion, which consists of $N$ spheres connected by $N-1$ extensible rods of negligible diameters [Fig.~\ref{fig:learning}(b)]. Each sphere has a radius $R$ and each rod has a length $\ell$ that can contract by a length $e$. We set $\ell = 10 R$ and $e=4 R$ in all cases in this study. An $N$-sphere system has a total of $2^{N-1}$ configurations, and each configuration can transition to $N-1$ different configurations by extending or contracting one of the connecting rods. Previous studies have used similar reconfigurable systems to generate net translation (e.g., Najafi-Golestanian's swimmer \cite{Najafi2004} and its variants \cite{Avron2005, Golestanian2009, Alouges2008}), rotation \cite{Dreyfus2005B}, and combined motion \cite{Earl2007, Alouges2013B}. Unlike the traditional approach where the swimming strokes were specified, here the spheres will self-learn propulsion policies based on knowledge gained by interacting with the surrounding medium via reinforcement learning \cite{Sutton1998}.

\subsection{Reinforcement learning} The use of reinforcement learning enables the swimmer to progressively learn how to act by interacting with the surrounding fluid [Fig.~\ref{fig:learning}(a)]. For a given configuration of the swimmer (the state, $s_n$) in the $n$-the learning step, the swimmer can extend or contract one of its rods (the action, $a_n$) to transforms from the current state to a new state. Such an action results in a certain displacement of the body centroid of the swimmer, which provides a reward ($r_n$) for the swimmer to measure the immediate success of the action relative to its goal in each learning step. We denote the body centroid of a swimmer as $\mathbf{c}_n=\sum_{i=1}^N \mathbf{x}_i(n)/N$, where $\mathbf{x}_i (n)$ represents the position vector of the $i$-th sphere. The transformation between states displaces $\mathbf{c}_n$ by $\Delta \mathbf{c}_n= \mathbf{c}_{n+1}-\mathbf{c}_n$. As we are interested in swimming motion along a desired direction $\hat{\mathbf{e}}$, we defined reward $r_n$ as the change of $\mathbf{c}_n$ due to the action: $r_n= \hat{\mathbf{e}}  \cdot \Delta \mathbf{c}_n$. The cumulative displacement of the body centroid in a learning process is given by $d=\sum_n \hat{\mathbf{e}} \cdot \Delta \mathbf{c}_n$. We scaled all lengths by the sphere radius $R$ in this work; hereafter, we use dimensionless cumulative displacement of the body centroid, $D = d/R$, to track the swimmer's location.

We implemented reinforcement learning based on the $Q$-learning algorithm \cite{Watkins92}, where the experience gained by the swimmer is stored in a $Q$-matrix, $Q(s_n,a_n)$. The matrix is an action-value function that captures the expected long-term reward for taking the action $a_n$ given the state $s_n$. Unlike model-based algorithms in reinforcement learning, $Q$-learning is a model-free algorithm that does not require a model for the environment and directly updates the action-value function. For its simplicity and expressiveness, we chose to use a classical $Q$-learning algorithm as a first example to elucidate the machine learning approach in designing low Re locomotion. After each learning step, $Q(s_n,a_n)$ is updated as
\begin{align} \label{eqn:Q}
Q(s_n, a_n) \leftarrow& \ Q(s_n, a_n) \\ \notag
&+\alpha\big[ r_n+\gamma \ \underset{a_{n+1}}{\mathrm{max}} Q(s_{n+1}, a_{n+1})-Q(s_n,a_n) \big]. \
\end{align}
Here, $\alpha$ is the learning rate ($0 \leq \alpha \leq 1$), which determines to what extent new information overrides old information in the $Q$-matrix. In a fully deterministic system, $\alpha=1$ gives the optimal convergence for the $Q$-matrix; hence, unless otherwise specified, we fixed $\alpha=1$. The $Q$-matrix encodes the adaptive decision-making intelligence of the swimmers by accounting for both immediate reward $r_n$ and maximum future reward at the next state, $\underset{a_{n+1}}{\mathrm{max}} Q(s_{n+1}, a_{n+1})$. The discount factor $\gamma$ assigns a weight to immediate versus future rewards  ($0\leq\gamma<1$). When $\gamma$ is small, the swimmer is shortsighted and tends to maximize the immediate reward; when $\gamma$ is large, the swimmer is farsighted and focuses more on the contribution of future rewards. 

To trade off exploitation of the gained knowledge and exploration of new solutions, we incorporated an $\epsilon$-greedy selection scheme: in each learning step, the swimmer chooses the best action advised by the $Q$-matrix with a probability $1-\epsilon$, or takes a random action with a probability $\epsilon$, which allows the swimmer to explore new solutions and avoid being limited to only locally optimal propulsion policies.

\subsection{Hydrodynamic interactions} The interaction between the spheres and the surrounding viscous fluid is governed by the Stokes equation, $\nabla p = \mu \nabla^2 \mathbf{v}$. For incompressible flows, $\nabla \cdot \mathbf{v}=0$. Here, $p$ and $\mathbf{v}$ represent, respectively, the pressure and velocity fields, and $\mu$ represents the dynamic viscosity. We used the Oseen tensor to consider the hydrodynamic interaction between spheres that are spaced far apart ($R/\ell \ll 1$) \cite{happel2012low, Najafi2004}. The linearity of the Stokes equation allows us to relate the velocities of the sphere $\mathbf{V}_j$ and the forces $\mathbf{F}_j$ acting on them as 
\begin{align}
\mathbf{V}_i = \sum_{j=1}^{N} \mathbf{H}_{ij} \mathbf{F}_j.
\end{align}
Here, the Oseen tensor $\mathbf{H}_{ij}$ for spheres is given by
\begin{align}
\mathbf{H}_{ij} =   \begin{cases}
\mathbf{I}/6\pi \mu R, & \text{if} \ i=j\\
(1/8\pi \mu |\mathbf{x}_{ij}|) (\mathbf{I}+\mathbf{x}_{ij} \mathbf{x}_{ij}/|\mathbf{x}_{ij}|^2), & \text{if} \ i\neq j
\end{cases},
\end{align}
where $\mathbf{I}$ is the identity matrix and $\mathbf{x}_{ij} = \mathbf{x}_j-\mathbf{x}_i$ denotes the vector between spheres $i$ and $j$. The instantaneous positions of the spheres $\mathbf{x}_i$ are determined by enforcing, respectively, the force-free and torque-free conditions
\begin{align}
\sum_{i=1}^N \mathbf{F}_i & =\mathbf{0}, \\
\sum_{i=1}^N \mathbf{F}_i \times \mathbf{x}_i &= \mathbf{0}.
\end{align}
The linearity and time-independence of the Stokes equation imply that the displacement of a swimmer depends only on the sequence of configurations (or states) changes of the swimmer. The sequence of state changes hence defines the propulsion policy of a low Re swimmer \cite{purcell1977life, lauga2009hydrodynamics}.

\begin{figure*}[t]
\centering
\includegraphics[width=0.65\textwidth]{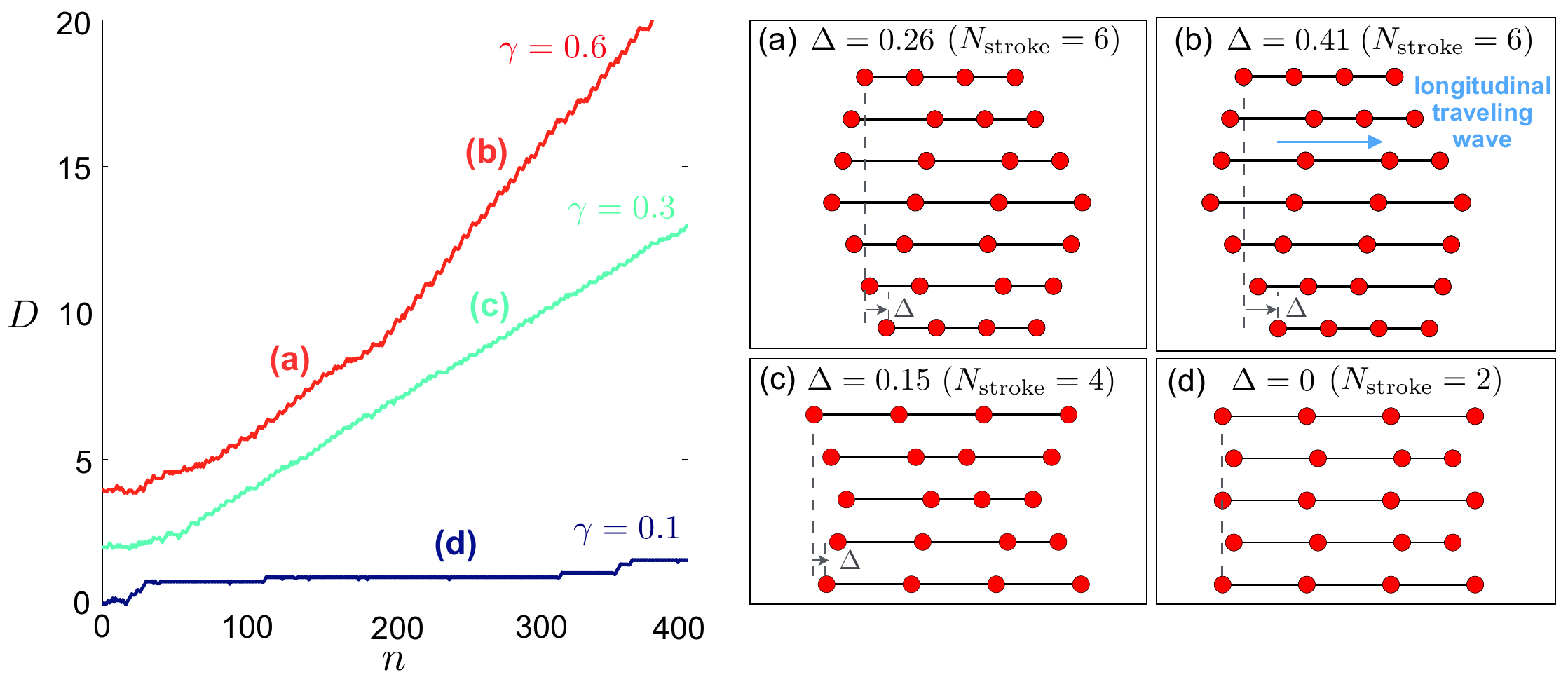}
\vspace{-0.3cm}
\caption{Exploring the parameters of reinforcement learning reveals how four-sphere swimmers progressively improve their propulsion policies. Left panel: The swimmers learn several propulsion policies depending on the discount factor $\gamma$. We set $\epsilon=0.05$. Panels (a)--(d) depict the swimming strokes of four different policies obtained throughout the learning process. These propulsion policies have varying number of strokes ($N_\text{stroke}$) and net displacement over one cycle ($\Delta$, scaled by $R$): (a) is the optimal policy with a longitudinal traveling wave pattern and (d) is a failed policy with which the swimmer does not swim.}
\label{fig:enhance}
\end{figure*}

\section{Results and Discussion}

\subsection{A self-learning three-sphere swimmer} We first considered a three-sphere swimmer ($N=3$), which has the minimal degrees of freedom for swimming at low Re \cite{purcell1977life,Najafi2004} [Fig.~\ref{fig:learning}(b)--(d), Movie S1]. The swimmer has four different configurations [Fig.~\ref{fig:learning}(b)]. In each learning step, the swimmer switches from one configuration to another, and updates the corresponding entry in the $Q$-matrix according to Eq.~\ref{eqn:Q}. Fig.~\ref{fig:learning}(c) depicts a typical episode of the self-learning process. The swimmer initially struggles to find a policy to swim forward and thus moves back and forth [left inset in Fig.~\ref{fig:learning}(c)], where $D$ remains close to $0$. The swimmer keeps exploring the surrounding medium by taking different actions and adapting its propulsion policy. After accumulating enough knowledge, the swimmer develops an effective propulsion policy that repeats the same sequence of action (except with $\epsilon$ probability, at which a random action is chosen), and swims with increasing $D$ [right inset in Fig.~\ref{fig:learning}(c)]. The propulsion policy obtained by our learning algorithm for a three-sphere swimmer is consistent with Najafi-Golestanian's swimmer \cite{Najafi2004}.

During the learning process, the entities in the $Q$-matrix are updated over the learning steps, and eventually converge to steady values [Fig.~\ref{fig:learning}(d)]. We observed that a swimmer starts repeating the same sequence of swimming strokes (when $n\approx 60$) well before all entities in the $Q$-matrix reach steady values [when $n>100$; see Fig.~\ref{fig:learning}(c) \& (d)]. The swimmer follows the same propulsion policy as long as the differences in entities in the $Q$-matrix remain on the same sign [Fig.~\ref{fig:learning}(d)]. 

The above example demonstrates, for the first time, how reinforcement learning enables a swimmer to self-learn how to swim with no prior knowledge on low Re locomotion.

\subsection{Extending to $N$-sphere swimmers} We extended this self-learning approach to systems with increased number of spheres. Unlike the three-sphere ($N=3$) swimmer, where only one propulsion policy leads to net translation, multiple propulsion policies are possible when the number of spheres increases \cite{Earl2007}. We first use a four-sphere system ($N=4$) as an example to discuss new features that can emerge in systems with more than three spheres ($N>3$); we then present results up to $N=10$ in Sec.~\ref{sec:brute}.

In Fig.~\ref{fig:enhance}, we use a four-sphere system to illustrate that a swimmer equipped with reinforcement learning not only can identify multiple propulsion policies [e.g., policies (a)--(c) in Fig.~\ref{fig:enhance}] but also can self-improve and evolve a better policy during the learning process [e.g., from policy (a) to (b) in Fig.~\ref{fig:enhance}, see Movie S2]. We consider three sets of simulations with different values of $\gamma$. First, with $\gamma=0.6$, the swimmer learns one propulsion policy [Fig.~\ref{fig:enhance}(a)] in the initial stage (Fig.~\ref{fig:enhance}, left panel). Through continuous learning, the swimmer keeps modifying its $Q$-matrix and eventually identifies a better propulsion policy [Fig.~\ref{fig:enhance}(b)], as indicated by the increased slope of $D$ in the left panel of Fig.~\ref{fig:enhance}. We note that this propulsion policy [Fig.~\ref{fig:enhance}(b)] is reminiscent of the propagation of a longitudinal traveling wave along a cell body \cite{Ehlers1996} and was shown to be optimal for a four-sphere system \cite{Earl2007}.

Continuous improvement of the propulsion policy does not always happen and depends on the choice of learning parameters, as demonstrated with $\gamma=0.3$ in Fig.~\ref{fig:enhance} (left panel). In this case, the swimmer learns a different but suboptimal propulsion policy [Fig.~\ref{fig:enhance}(c)] and cannot improve the policy in the subsequent learning process. When $\gamma$ is even lower (e.g., $\gamma=0.1$), the system fails to learn any effective propulsion policy; the policy harvested corresponds to a back-and-forth motion that leads to zero net translation [Fig.~\ref{fig:enhance}(d)]. We note that even though the swimmer with $\gamma=0.1$ did not learn to swim, it displayed a slight drifting biased towards the positive direction. This resulted from the combined effect of the $\epsilon$-greedy exploration steps, which allow the swimmer to act against the advice of the $Q$-matrix, and the encouragement of overall positive displacement in the learning process by the rewards. The complex dependences of propulsion policy on the learning parameters motivate the parametric studies detailed below.

\subsubsection{Influences of learning parameters on swimmers} Here, we systematically investigated the influences of learning parameters on a four-sphere swimmer. The learning outcome depends on how much the swimmer values an immediate reward ($\gamma$), how often the swimmer explores randomly ($\epsilon$), and how many learning steps the swimmer takes ($N_\text{learn}$).

\begin{figure*}[t]
\centering
\includegraphics[width=1\textwidth]{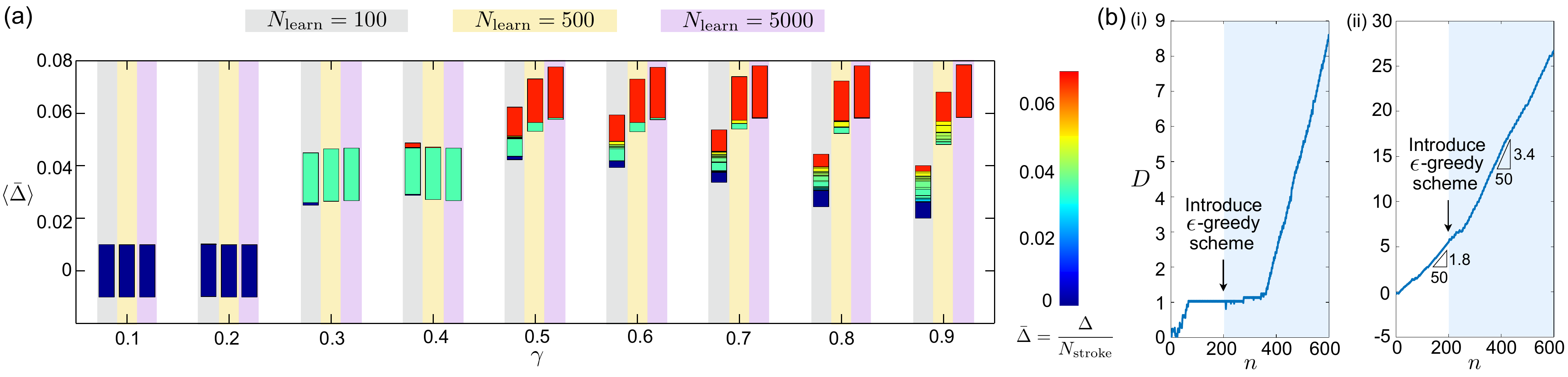}
\vspace{-0.3cm}
\caption{Effects of the learning parameters and the trade-off between exploration and exploration on the learning outcome. (a) The learning outcome as a function of $\gamma$ and total number of learning steps $N_\text{learn}$. Here, we varied $\gamma$ from 0.1--0.9 in increments of 0.1 and $N_\text{learn}$ between 100 (shaded regions in grey), 500 (yellow) and 5,000 (purple), with a fixed $\epsilon=0.1$. Each bar represents the results of 1,000 individual simulations and the colors of each segment represent the net displacement over one cycle divided by the number of strokes, $\bar{\Delta}$. Although distinct propulsion policies can have the same $\bar{\Delta}$ and, hence, the same color, they are separated by a black border in the bar. The middle of each bar represents $\langle \bar{\Delta} \rangle$ (the averaged $\bar{\Delta}$) for the 1,000 simulations. (b) Introducing an $\epsilon$-greedy scheme with $\epsilon=0.05$ allows the swimmer to escape from locally trapped policies and evolve a better propulsion policy: (i) The swimmer transitions from a failed to a successful policy at $\gamma=0.4$. (ii) The swimmer transitions from a sub-optimal to an optimal propulsion policy at $\gamma=0.9$.}
\label{fig:parameters}
\end{figure*}

We explored different possibilities by performing Monte-Carlo-type simulations with random initial states, where we fixed $\epsilon=0.1$ and varied $\gamma$ and $N_\text{learn}$. We performed $1,000$ simulations for each set of parameters and extracted the resulting propulsion policies given by the $Q$-matrix after learning. In order to distinguish propulsion policies with different numbers of strokes per cycle, we characterized each propulsion policy by the displacement per stroke, $\bar{\Delta} = \Delta/N_\text{stroke}$, which divides the net displacement over one cycle $\Delta$ by the number of strokes in the cycle $N_\text{stroke}$. The simulation results revealed three main findings [Fig.~\ref{fig:enhance}(b)]:

(1) Given sufficiently large $N_\text{learn}$ (e.g., $N_\text{learn}=5000$), the learning process evolves into three major outcomes, depending on the value of $\gamma$ [Fig.~\ref{fig:parameters}(a)]. At a small $\gamma$ ($ \le 0.2$), the swimmer fails to swim [a two-stroke policy in Fig.~\ref{fig:enhance}(d)]. At an intermediate $\gamma$ (e.g., $\gamma=0.3-0.4$), the swimmer identifies an effective but suboptimal policy [a four-stroke policy in Fig.~\ref{fig:enhance}(c)]. At a large $\gamma$ ($\ge0.5$), the swimmer learns the optimal policy [a six-stroke policy in in Fig.~\ref{fig:enhance}(b)], which corresponds to a longitudinal traveling wave pattern.

(2) There exists a threshold of $\gamma$, below which the swimmer cannot learn the optimal propulsion policy (e.g., for a four-sphere system, the critical $\gamma \lesssim 0.5$). The learning process leads to suboptimal policies even with many learning steps. This occurs because compared to the suboptimal policies, the optimal policy involves more swimming strokes, including those that contribute immediate, negative rewards in the cycle. Therefore, only a far-sighted swimmer (large $\gamma$) can learn the optimal propulsion policy. Before the propulsion policy converges (e.g., when $N_\text{learn}=100$), a small portion of swimmers  at $\gamma=0.4$ follow the optimal policy due to random initialization of the $Q$-matrix, but the policy eventually  converges to a suboptimal policy at large $N_\text{learn}$ (e.g., $N_\text{learn}=5,000$).

(3) For a given number of learning steps $N_\text{learn}$, an optimal $\gamma$ maximizes the portion of swimmers that can acquire the optimal propulsion policy [Fig.~\ref{fig:parameters}(a)]. As $N_\text{learn}$ increases, the optimal $\gamma$ increases its value from $\gamma \approx 0.5$ for $N_\text{learn}=100$ to $\gamma \approx 0.7$ for $N_\text{learn}=500$. These results illustrate that while a sufficiently large $\gamma$ is necessary to acquire the optimal policy, an excessively large $\gamma$ (e.g., $\gamma=0.9$) can delay the learning of the optimal policy because the swimmer becomes too far-sighted and largely ignores immediate rewards for new possibilities. In addition, when there is only a small number of learning steps [e.g., $N_\text{learn}=100$ in Fig.~\ref{fig:parameters}(a)], this emphasis on long-term benefits results in harvesting more distinct policies as $\gamma$ increases, which can hamper the overall learning outcomes [see decrease in $\bar{D}$ for learning processes with $N_\text{learn}=100$ and $\gamma>0.5$ in Fig.~\ref{fig:parameters}(a)].

The effects of $\epsilon$ on self-learning the propulsion policy are obvious when we compared $\epsilon=0$ and $\epsilon>0$ [Fig.~\ref{fig:parameters}(b)]. When $\epsilon=0$ (greedy policy), the swimmer can get trapped in certain suboptimal propulsion policies. For instance, a swimmer may be trapped in a failed policy that yields no net displacement [white region in Fig.~\ref{fig:parameters}(b)-i] or an effective but suboptimal policy [white region in Fig.~\ref{fig:parameters}(b)-ii]. In either case, the introduction of $\epsilon>0$ (epsilon-greedy policy) helps kick the swimmer away from these locally trapped policies, thus enabling the swimmer to continually improve its propulsion policy to its fullest extent for a given value of $\gamma$ [blue regions in Fig.~\ref{fig:parameters}(b)-i \& ii].

Taken together, these findings reveal how learning parameters influence the robustness of the self-learning approach for identifying effective propulsion policies.

\begin{figure}[t]
\centering
\includegraphics[width=0.33\textwidth]{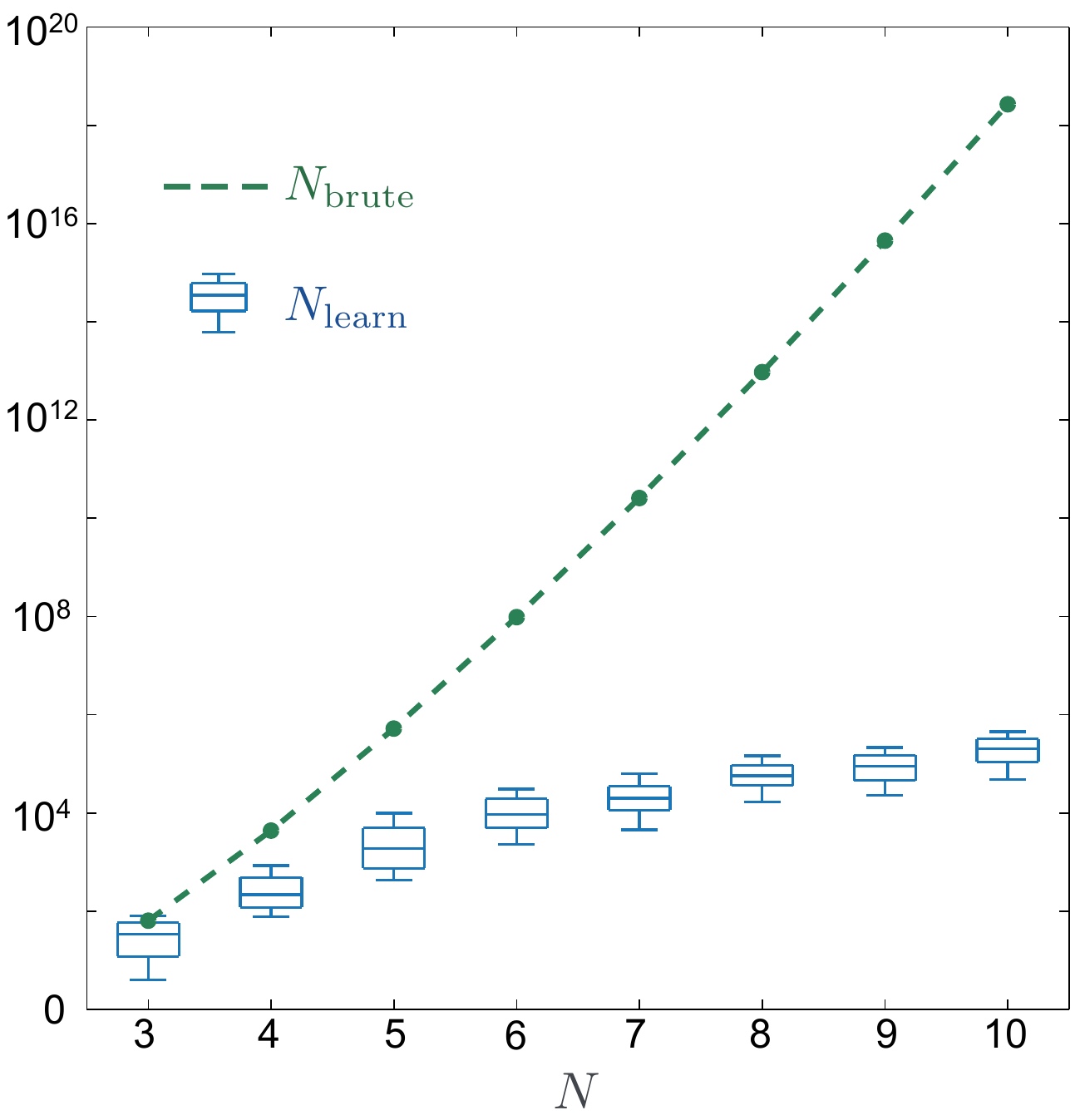}
\vspace{-0.3cm}
\caption{The number of learning steps $N_\text{learn}$ required for an $N$-sphere swimmer to learn the traveling-wave propulsion policy is shown in the box plot: the mid lines represent the median, and the box's upper and lower bounds indicate the interquartile range; the upper and lower whiskers denote the 9th and 91st percentile of the simulation data, respectively. The vertical scale is logarithmic in base 10. We performed $100$ simulations for each value of $N$, where $\gamma=0.9$ and $\epsilon=0.1$. As a comparison, a brute-force search through all combinations with the same number of strokes as the traveling-wave policy requires $N_\text{brute}=2(N-1)^{2N-1}$ number of simulation steps (green dashed line).}
\label{fig:performance}
\end{figure}

\subsubsection{Systems with increased number of spheres} \label{sec:brute} 
Next we probed the performance of this simple $Q$-learning approach for systems with increased degrees of freedom. We considered self-learning swimmers consisting of up to ten spheres (i.e., $N=3$ to $N=10$). With sufficiently large number of learning steps, we found that these $N$-sphere swimmers all learn the propulsion policy reminiscent of the longitudinal traveling wave pattern [e.g., Fig.~\ref{fig:enhance}(b)]. For every $N$ considered, we performed 100 simulations and evaluated the minimum number of learning step required for a swimmer to learn the traveling-wave propulsion policy. We display the required learning steps $N_\text{learn}$ as a function of $N$ in Fig.~\ref{fig:performance} (box plots). More learning steps are required for systems with increased number of spheres as expected, but the rate of increase levels off for larger values of $N$. As a remark, the threshold for $\gamma$ to learn such traveling-wave policies increases with $N$, because a swimmer needs to become more far-sighted in order to learn a policy involving increasingly more strokes as $N$ increases. We set $\gamma=0.9$ for all cases considered to ensure the occurrence of these traveling-wave policies.

\begin{figure}[t!]
\centering
\includegraphics[width=0.33\textwidth]{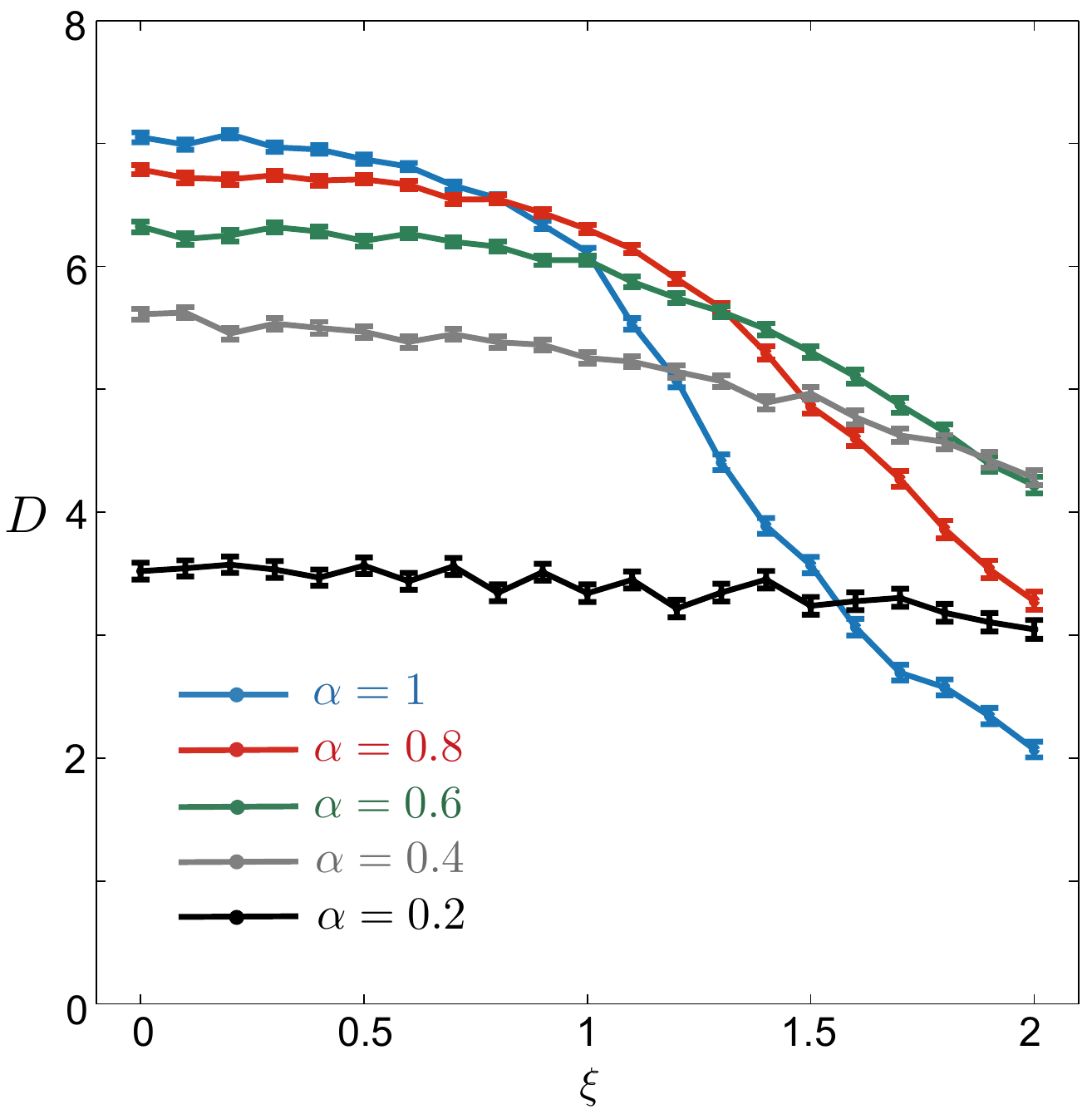}
\vspace{-0.3cm}
\caption{A self-learning swimmer can propel robustly in a noisy environment. We considered a three-sphere swimmer and measured its cumulative displacement $D$ after $N_\text{learn}=200$ under increasing noise level $\xi$ and various learning rates $\alpha$. Each data point represents the mean of $D$ for 1,000 simulations. The error bars denote the standard error of the mean. In all cases, $\gamma=0.7$.}
\label{fig:noise}
\end{figure}

We note that the learning algorithm harvests effective policies from a pool consisting of a tremendous number of stroke combinations as the number of spheres increases. Here we provide a crude estimate of the size of the pool of combinations. For an $N$-sphere swimmer, the traveling wave policy harvested after the learning steps shown in Fig.~\ref{fig:performance} has $2(N-1)$ number of strokes. The number of combinations that have the same number of strokes is hence  $(N-1)^{2(N-1)}$ because the swimmer has $N-1$ choices of which rod to actuate for each stroke. If one were to perform a brute-force search, a total number of $N_\text{brute}=2(N-1)^{2N-1}$ steps are required to go through all combinations (green dashed line, Fig.~\ref{fig:performance}). The brute-force search becomes quickly intractable as $N$ increases and the number of simulations steps required are orders of magnitude greater than the required learning steps (box plots). We remark that while the classical $Q$-learning algorithm employed here can be extended to consider even larger values of $N$, there exists a vast potential for improving the scalability for large state and action spaces using more advanced machine learning approaches \cite{Mnih15}. A search for the optimal machine learning algorithm is beyond the scope of this work. Here we take the first step to quantify how much a simple learning algorithm can already perform better than with brute-force search for more complex systems.

\begin{figure*}[t!]
\centering
\includegraphics[width=0.65\textwidth]{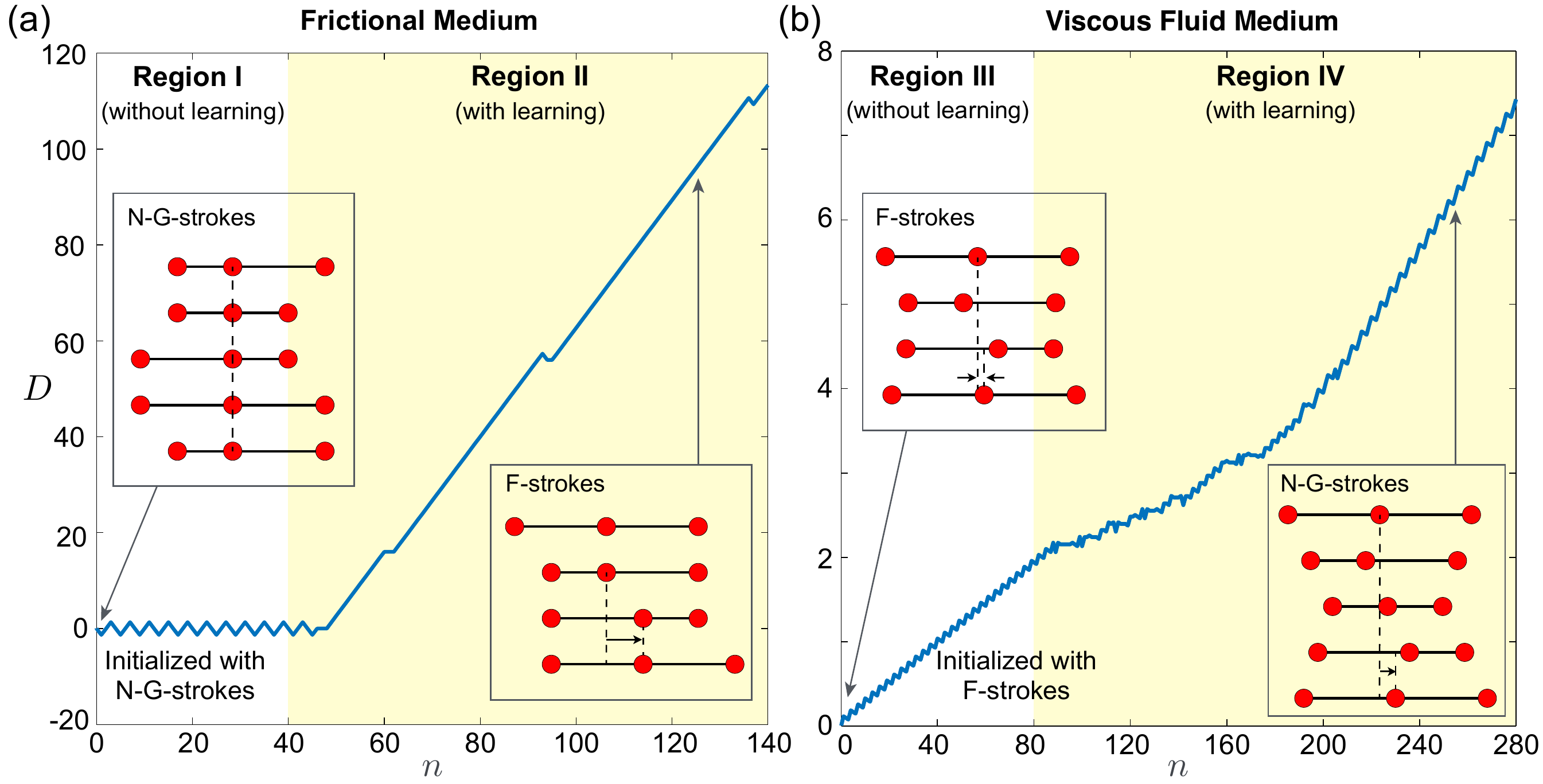}
\caption{A self-learner adapts its locomotory gait to propel effectively in vastly different media. (a) Although Najafi-Golestanian's strokes (termed N-G-stokes; left inset) are optimal for propulsion in a viscous fluid, performing the N-G-strokes initially (from $n=0-40$; region I) in a frictional medium without reinforcement learning leads to back-and-forth motion with no net displacement $D$. When the learning algorithm is turned on, the self-learner rapidly identifies new locomotory gaits (termed F-strokes; right inset) to propel effectively in the frictional medium. (b) A swimmer initialized with F-strokes (left inset) also propels in a viscous fluid medium (from $n=0-80$; region III; without learning). Nevertheless, with reinforcement learning (region IV), the self-learning swimmer can explore, relearn, and evolve to the N-G-strokes (right inset), which remain the optimal propulsion policy in a viscous fluid even when both rods are allowed to actuate simultaneously. In (a) \& (b), $\gamma=0.7$, $\epsilon=0.05$.} 
\label{fig:adaptive}
\end{figure*}

\subsection{A self-learning swimmer under noise} We assessed how a self-learning swimmer behaves under the influence of random noises from the environment. In each learning step, we introduced noise to the displacement (and hence reward) of the swimmer: $r_n= \hat{\mathbf{e}}  \cdot \Delta \mathbf{c}_n (1+\xi U)$, where $\xi$ is the noise level and $U$ is a random variable with a uniform distribution in $[-1,1]$. To illustrate, we considered the case of a three-sphere swimmer and used cumulative displacement $D$ at $N_\text{learn}=200$ as a metric for the performance of the self-learning swimmer under increasing noise levels $\xi$ (Fig.~\ref{fig:noise}, Movie S3). When the noise is weak ($\xi \leq 0.5$), the swimmer with a learning rate $\alpha =1 $ (blue solid line) performs the best. As the noise level increases, $\alpha=1$ no longer guarantees the best performance and different optimal values of $\alpha$ exist depending on the noise level. Remarkably, when the noise level is as large as 100\% of the instantaneous displacement ($\xi=1$), a self-learning swimmer with $\alpha=0.8$ still reaches over 85\% of the mean displacement compared with the noise-free environment ($\xi=0$); even when the noise level is twice as much as the instantaneous displacement ($\xi=2$), a swimmer with a reduced learning rate in the range of $\alpha=0.4-0.6$ can retain over 60\% of its noise-free performance (Fig.~\ref{fig:noise}). Thus, a self-learning swimmer can robustly adapt to swim in a noisy environment, and further improve its performance by adjusting its learning rate.

\subsection{Adaptive locomotion across different media} \ Finally, we demonstrated the adaptivity of this self-learning approach in a medium vastly different from viscous fluids -- a frictional environment \cite{Guo2008, Hu2009, Maladen2009} where motion arises by the interaction of surface friction $\mathbf{F}_i$ and net driving forces $\mathbf{f}_i$ exerted on each sphere by the rods. We again considered a three-sphere swimmer, but allowed both rods to move in the same step, thereby enabling free transitions between the four states in Fig.~\ref{fig:learning}B. We restricted our analysis to a standard Coulomb sliding friction law \cite{Guo2008, Hu2009}: when the magnitude of the net driving force on a sphere is greater than the sliding friction $F_s$ (i.e., $|\mathbf{f}_i|>F_s$), the friction acting on the sphere is given by $\mathbf{F}_i = - F_s \hat{\mathbf{V}}_i$, where $\hat{\mathbf{V}}_i= \mathbf{V}_i/|\mathbf{V}_i|$ is the velocity direction of the sphere. When $|\mathbf{f}_i|\le F_s$, the static friction balances the net driving force, $\mathbf{F}_i = - \mathbf{f}_i$.      

In the low Froude number (Fr) limit, inertial forces are subdominant to frictional forces. Frictional forces are therefore transduced directly to velocities instead of accelerations, similar to the low Re regime in viscous fluid flows \cite{Hu2009}. As a result, locomotion in frictional media in the $\text{Fr}=0$ limit is kinematic (also a feature of Stokesian locomotion), in that the net displacement is independent of its rate but only the sequence of deformations \cite{Hatton2013}. Despite the similarities between frictional and viscous fluid media, a key difference is the absence of hydrodynamic interactions in the frictional medium. This difference renders Najafi-Golestanian's strokes (N-G-strokes) of a three-sphere swimmer ineffective in frictional media (region I in Fig.~\ref{fig:adaptive}; without learning, Movie S4). Nevertheless, when we turn on reinforcement learning and allow simultaneous actuation of both rods, the self-learner rapidly adapts to the frictional medium and identifies a new, effective propulsion policy (region II; F-strokes). We note that it takes significantly fewer steps to learn propulsion policies in the frictional medium than in the viscous medium because the F-strokes do not involve steps that contribute intermediate, negative rewards. Finally, we found that the new locomotory gaits identified in frictional media (F-strokes) also propel a swimmer in a viscous fluid (region III; without learning, Movie S5); nevertheless, a swimmer with reinforcement learning will explore, re-learn, and evolve a better propulsion policy to adapt to the surrounding medium (region IV; returning to the N-G-Strokes). The adaptivity demonstrated here a first step in realizing a smart ``amphibian'' micro-robot that can move effectively in both liquid and solid terrains by adjusting its locomotory gait.

\section{Concluding remarks} 
The design of successful locomotory gaits for micro-robots is subject to both constraints by physical laws at microscales and uncontrolled environmental factors in biological media. Designing micro-robots to traverse complex biological environments with vastly varying and often-unknown properties remains an unresolved challenge. Diverging from the traditional paradigm of low Re locomotion, where the locomotory gaits are specified \textit{a priori}, here we present a machine learning approach that allows a robot to self-learn effective propulsion policies based on its interactions with the surrounding environment. Resulting self-learning robots can identify effective propulsion policies in a viscous fluid, propel robustly in a noisy environment, and adapt their locomotory gait to move in a frictional medium. The demonstrated adaptivity circumvents unpredictability that can arise in complex environments. This reinforcement learning approach to low Re locomotion applies as well to other types of swimmers that have well-defined states; here we illustrate key features of a self-learning swimmer through a $N$-sphere swimmer only as a minimal example.

These initial steps spark future works in several directions. First, the theoretical model studied here is amenable to future experimental implementation. For instance, colloidal particles actuated by selective contraction of photoactive soft materials under dynamic light fields \cite{Palagi16} combined with a real-time microscopy system for position tracking \cite{Landin} provide a viable experimental platform to implement self-learning $N$-sphere swimmers. Second, we demonstrated in this work directed translation as an example but the same self-learning approach applies to directed rotation and hence more complex maneuvers. 
Third, we chose a classical $Q$-learning algorithm here for its simplicity and expressiveness. While we demonstrated its effectiveness over brute-force search for swimmers consisting of up to ten spheres, the pursuit of more scalable machine learning approaches when the system becomes increasingly complex is an important next step \cite{Mnih15}. This self-learning approach opens an alternative avenue to designing the next generation of smart micro-robots with robust locomotive capabilities. 

%

\end{document}